\magnification \magstep1
\input amssym.def
\input amssym.tex
\bigskip
\centerline{\bf Relic Radiation from an Evaporating Black Hole}
\bigskip
\centerline{A.N.St.J.Farley and P.D.D'Eath}
\bigskip
\bigskip
\centerline{Department of Applied Mathematics and Theoretical Physics,
Centre for Mathematical Sciences,} 
\smallskip
\centerline{University of Cambridge, Wilberforce Road, Cambridge CB3
0WA, United Kingdom} 
\bigskip
\centerline{Abstract}
\bigskip
A non--string--theoretic calculation 
is presented of the microcanonical entropy 
of relic integer--spin Hawking radiation 
--- at fixed total energy 
$E$ 
--- from an evanescent, 
neutral, 
non--rotating four--dimensional black hole. 
The only conserved macroscopic quantity is the total energy 
$E$ 
which, 
for a black hole that evaporates completely, 
is the total energy of the relic radiation. 
Through a boundary--value approach, 
in which data for massless, 
integer--spin perturbations 
are set on initial and final space--like hypersurfaces, 
the statistical--mechanics problem becomes, 
in effect, 
a one--dimensional problem,
with the `volume' of the system determined by the real part 
of the time separation at spatial infinity 
-- the variable conjugate to the total energy. 
We count the number of field configurations 
on the final space--like hypersurface that have total energy 
$E$, 
assuming that initial perturbations are weak. 
We find that the density of states 
resembles the well--known Cardy formula. 
The Bekenstein--Hawking entropy is recovered if the real part 
of the asymptotic time separation is of the order 
of the semi--classical black-hole life--time. 
We thereby obtain a statistical interpretation of black--hole entropy. 
Corrections to the microcanonical entropy are computed, 
and we find agreement with other approaches in terms 
of a logarithmic correction to the black--hole area law, 
which is 
{\it universal} 
(independent of black--hole parameters).  
This result depends crucially upon the discreteness 
of the energy levels. 
We dicuss the similarities of our approach with the transition 
from the black--hole to the fundamental--string r\'egime 
in the final stages of black--hole evaporation.  
In addition, 
we find that the squared coupling, 
$g^{2}$, 
which regulates the transition from a black hole 
to a highly--excited string state, 
and 
{\it vice versa}, 
can be related to the angle, 
$\delta$, 
in the complex--time plane, 
through which we continue analytically 
the time separation at spatial infinity. 
Thus, 
in this scenario, 
the strong--coupling r\'egime corresponds to a Euclidean black hole, 
while the physical limit of a Lorentzian space--time 
(the limit as 
$\delta\to{0}_{+}$) 
corresponds to the weak--coupling r\'egime. 
This resembles the transition of a black hole to a highly--excited 
string--like state which subsequently decays into massless particles, 
thereby avoiding the naked singularity.\par
\bigskip
\noindent 
{\bf 1. Introduction }
\medskip
\indent
Bekenstein [1] was the first to associate an intrinsic entropy 
with a black hole, 
such that black-hole entropy measures one's ignorance 
about the black hole's internal state. 
The logarithm of the number of distinct internal configurations 
of the hole, 
when one studies the microcanonical ensemble, 
in which the macroscopic external parameters 
(mass, 
charge and angular momentum) 
are fixed, 
then determines the microcanonical entropy. 
These internal configurations are related 
to different possible pre--collapse configurations, 
which result in the formation of a stationary black hole 
with the same external parameters. 
In this interpretation, 
the black--hole entropy is the logarithm of the number 
of distinct ways in which the black hole 
may be formed by infalling matter.

In relating the dynamical degrees of freedom of a black hole 
to its entropy, 
it was suggested that these degrees of freedom refer to states of 
{\it all} 
fields which are located inside the black hole [2]. 
In particular, 
modes of fields inside the black hole in the vicinity 
of the horizon give the dominant contribution to the entropy. 
Averaging over the states located outside the hole then enables 
the statistical--mechanical entropy of the black hole to be determined.

The main aim of our approach is to give the black--hole entropy 
a statistical--mechanical interpretation in terms of a counting of states. 
We proceed by considering a 
(quantum or classical) 
boundary--value problem, 
in which data for massless, 
integer--spin, 
wave--like perturbations are set on initial and final
asymptotically--flat space--like hypersurfaces 
$\Sigma_{I}$ 
and 
$\Sigma_{F}{\,}$, 
separated by a large proper time 
$T$ 
at infinity [3]. 
(Fermions will be considered in another paper.) 
In [3], 
this led to calculations of the quantum amplitude 
(not just the probability) 
to go from 
(say) 
an initial spherically--symmetric pre--collapse configuration 
to a prescribed final distribution of radiative fields 
after the black hole has evaporated completely.  
In order to make the classical boundary--value problem well--posed, 
one must rotate the asymptotic proper time 
$T$ 
into the lower complex plane; 
the real--$T$ 
hyperbolic boundary--value problem for wave--like field perturbations 
is not well--posed [4].

Below, 
it will become clear that the real part of the asymptotic proper time 
$T$ 
between the initial and final hypersurfaces 
plays a major part in the computation of the entropy. 
Apart from the total ADM energy 
--- the only macroscopic quantity appearing which encodes 
the three--dimensional geometry of the initial and final hypersurfaces 
--- the only other macroscopic quantity 
is the one-dimensional `volume' of the system, 
here effectively given by 
$T{\,}$.  
The 'volume', 
being conjugate to the fixed total energy at infinity,
assumes a large range of values. 
However, 
we shall not take the thermodynamic or infinite--volume limit. 
We show that, 
when the `volume' 
$T$ 
is of the order of the evaporation time of the black hole, 
and when the given 
(fixed) 
total energy,
measured after complete evaporation of the black hole, 
is 
$E{\,}$,
then the logarithm of the number of final radiation microstates 
is given by the usual expression for the initial entropy 
of the black hole, 
which semi--classically has the form 
${\,}4{\pi}{\,}k_{B}{\,}M^{2}{\,}(m_{p})^{-2}{\;}$.
This gives a statistical--mechanical interpretation 
for the black--hole entropy.

The final state on the final space--like hypersurface 
$\Sigma_{F}$ 
is labelled by a set of variables 
$\{E,b_{j}\}{\,}$.  
Here, 
the 
$\{b_{j}\}{\,}$ 
denote the massless, 
integer--spin 'Fourier' amplitudes 
[see Eq.(1.6) below], 
$j$ 
labels the quantum numbers associated with the relic radiation, 
and 
$E$ 
is the fixed total energy of the relic radiation. 
For simplicity, 
consider the example of small perturbations 
${\,}\varepsilon\phi^{(1)}{\,}$, 
about a spherically--symmetric background 
$\Phi(t,r){\,}$, 
of a massless real scalar field
${\,}\phi{\,}
={\,}\Phi{\,}+{\,}\varepsilon{\,}\phi^{(1)}{\,}$, 
where
$\varepsilon$
is a small expansion parameter. 
One expands the perturbation function 
$\phi^{(1)}$ 
in the form [3]
$$\phi^{(1)}(t,r,\theta,\phi){\quad} 
={\quad}{{1}\over{r}}{\;}{\,}\sum^{\infty}_{\ell = 0}{\;} 
\sum ^{\ell}_{m=-\ell}{\;} Y_{\ell m}(\Omega){\;}{\,}R_{\ell m}(t,r)
{\quad}.
\eqno(1.1)$$
\noindent 
The background metric is taken in the form 
$$ds^{2}{\quad} 
={\quad}-{\;}{\,}e^{b(t,r)}{\,}dt^{2}{\;}{\,} 
+{\;}{\,}e^{a(t,r)}{\,}dr^{2}{\;}{\,} 
+{\;}{\,}r^{2}{\;}
\bigl(d{\theta}^{2}{\,}+{\;}{\sin}^{2}\theta{\;}d{\phi}^{2}{\,}\bigr)
{\quad}.
\eqno(1.2)$$
\noindent
The perturbed scalar wave equation at late times leads to the 
$(\ell{\,},m)$ 
mode equation 
$$\Bigl(e^{(b-a)/2}\partial_{r}\Bigr)^{2}R_{\ell m}{\,} 
-{\;}\bigl(\partial_{t}\bigr)^{2}R_{\ell m}{\,} 
-{\,}{{1}\over{2}}{\,}
\Bigl(\partial_{t} (a-b)\Bigr)\Bigl(\partial_{t}R_{\ell m}\Bigr){\,} 
-{\;}V_{\ell}(t,r){\,}R_{\ell m}{\quad} 
={\quad}0
{\;},
\eqno(1.3)$$
\noindent
where 
$$V_{\ell}(t,r){\quad} 
={\quad}{{e^{b(t,r)}\over{r^{2}}}}{\;} 
\biggl(\ell(\ell +1){\;}+{\;}{{2m(t,r)}\over{r}}\biggr)
{\quad}.
\eqno(1.4)$$
\noindent
Here, 
$m(t{\,},r)$ 
is defined by 
$$\exp\Bigl(-{\;}a(t,r)\Bigr){\quad}{\;} 
={\quad}{\;}1{\;}{\,}-{\;}{\,}{{2m(t,r)}\over{r}}
{\quad}.
\eqno(1.5)$$
\indent 
Near the final hypersurface 
$\Sigma_{F}{\,}$, 
{~}one {~}can {~}decompose {~}the {~}radial {~}wave {~}functions 
$R_{\ell m}(t,r)$ 
harmonically with respect to time,
and the resulting functions 
$R_{k\ell}(r)$ 
can be normalised in a convenient way [3].  
With this normalisation, 
when one works near the final hypersurface 
$\Sigma_{F}{\,}$, 
one can write 
$$\phi^{(1)}{\quad} 
={\quad}{{1}\over{r}}{\;}{\,}\sum_{\ell m}{\;}\int_{0}^{\infty}
dk{\;}{\;}b_{k\ell m}{\;}{\,}R_{k\ell}(r){\;}{\,}Y_{\ell m}(\Omega)
{\quad}.
\eqno(1.6)$$
\noindent
This defines the 
${\,}b_{j}{\,}$ 
coefficients for spin 
$s{\,}={\,}0{\,}$;  
thus, 
$j$ 
denotes the collective indices 
$k{\ell}m{\,}$.
The condition that 
$E{\,}$ 
is fixed corresponds physically to a non--equilibrium situation, 
where there is an exchange of energy between the black hole and infinity, 
which results 
(after a long time--interval)
in the disappearance of the black hole 
(and possible later re--appearance, 
etc.). 
As there are no other conserved quantities apart from the total energy, 
an isolated system in equilibrium, 
having total energy 
$E{\,}$, 
would sample all its eigenstates at that energy with equal probability. 
The amplitudes 
$\{b_{j}\}{\,}$, 
however, 
correspond to different final field/particle configurations 
(microstates) 
but identical total energy 
$E$ 
(macrostate).  
The coarse--graining arises from the loss of phase information 
about the final configuration. 
The microcanonical entropy, 
therefore, 
is the logarithm of the number of different microstates 
that correspond to the same macrostate.

One would also like to understand in more detail what happens 
in the final stages of black--hole evaporation.  
For example, 
it is expected that, 
within a semi--classical and purely Lorentzian picture, 
a naked singularity would form just as the black hole finally disappears.  
Were the black hole to disappear completely, 
then there would be a breakdown in the unitary evolution 
of quantum mechanics [5]. 
That is, 
pure quantum states would evolve into mixed states 
when the information about black--hole states located 
inside the horizon is lost. 
Here, 
though, 
we follow the approach of [3], 
in which the proper time 
$T$ 
at infinity is displaced into the lower complex half--plane, 
and so any singularity in the classical solution is avoided.  
This leads us to consider the microcanonical ensemble;  
pure quantum states evolve into pure quantum states,
and there is no information loss.  
In another paper, 
we give in more detail the resulting unitary description 
of quantum black--hole evolution [6].

In Sec.2, 
we compute the total energy of the relic Hawking flux 
crossing the final hypersurface, 
in the limit of a large total energy or mass. 
In Sec.3, 
we calculate the microcanonical density of states as a function 
of total energy and `volume', 
that is, 
of the asymptotic--time interval 
$T$.  
In Sec.4, 
we summarise the theory of the black hole/fundamental 
string transition in the final stages of evaporation, 
in order to exhibit the similarities with our approach{~~} 
--- {~}in a recent paper [7],
a cosmological analogue of the black hole/string transition
was described.
The idea is that massive elementary particles 
behave like black holes at large coupling. 
In string theory, 
there are very massive states, 
and large numbers of states of a given mass.  
In this approach, 
the large degeneracy is regarded as the origin 
of the Bekenstein--Hawking entropy. 
There follows a discussion of the thermodynamics in Sec.5. 
Sec.5.1 is devoted to deriving the semi--classical Bekenstein--Hawking 
(B--H) 
entropy formula through constraints on the one--dimensional volume. 
In Sec.5.2, 
we compute the corrections, 
beyond the semi--classical order,
to our microcanonical entropy formula, 
finding agreement with other approaches as to the 
{\it universal} 
negative leading--order correction. 
Sec.6 puts some of the ideas of Secs.3 and 5 on a firmer basis,
in terms of the black hole/string transition theory of Sec.4.  
Here, 
we use the probabilistic interpretation of the theory, 
derived from the quantum amplitudes associated with the boundary data, 
and make contact with the Euclidean approach 
to black--hole evaporation.\par
\bigskip
\noindent 
{\bf 2. Energy Constraint}
\medskip
\indent 
The total energy, 
$E$, 
of the radiation fields on the final hypersurface 
$\Sigma_{F}{\,}$, 
is given by
$$E{\quad}
={\quad}\int_{\Sigma_{F}}d^{3}x{\;}{\;}\sqrt{-{\,}\gamma}{\;}{\;}{\cal{H}}
{\quad},
\eqno(2.1)$$
\noindent
where 
${\cal{H}}=e^{-b}{\,}T^{(2)}_{~~tt}{\,}$, 
with 
$T^{(2)}_{~~\mu\nu}$
being the lowest--order perturbed energy--momentum tensor, 
which is quadratic in the field perturbations.
There is no background matter contribution, 
in view of the final boundary condition that only relic 
Hawking radiation is present at late times. 
The gravitational background geometry is described by a metric 
$\gamma_{\mu\nu}{\,}$, 
which is spherically symmetric, 
but time--dependent,
being of the form given in Eq.(1.2). 
The background approximation is described more fully in [3]. 
At late times, 
the background metric 
can be accurately described by a Vaidya metric [8]. 
To simplify the exposition, 
since, 
in principle, 
fields of different spins 
$s{\,}$
may be present, 
we shall consider here only a massless spin--0 perturbation, 
writing the perturbative scalar field as
$A^{(1)}(x)$; 
the further inclusion of spin--1 and spin--2 perturbations 
is straightforward [3].  
On integrating by parts, 
using the linearised massless scalar--field equation 
and employing the {~}boundary {~}condition {~}of {~}regularity 
{~}at {~}the {~}origin,
$\{r=0\}$, 
{~}together {~}with {~}use {~}of {~}the {~}asymptotic {~}behaviour
$r^{2}A^{(1)}{\,}\bigl(\partial_{r}A^{(1)}\bigr){\;}{\;}
\sim{\;}{\;}r^{-1}{\;}{\,}$ 
(or faster) 
at large 
$r{\,}$, 
we find, 
for our perturbative scalar field:
$$\eqalign{&E{\quad} 
={\quad}
{{1}\over{2}}{\;}
{\int}{\,}d\Omega\int_{\Sigma_{F}}dr{\;}{\,}r^{2}{\;}{\,}
e^{{{1}\over{2}}(a-b)}{\;}
\Bigl(\dot{A}^{(1){\,}2}-{\,}A^{(1)}\ddot{A}^{(1)}\Bigr)\cr
&{\qquad}{\qquad}-{\;}{\;}{{1}\over{2}}{\;}{\int}{\,}d\Omega{\,}
\int_{\Sigma_{F}}dr{\;}{\,}r^{2}{\;}{\,}
e^{{{1}\over{2}}(a-b)}{\;}
A^{(1)}{\,}\dot{A}^{(1)}{\;}{{1}\over{2}}{\;} 
\bigl(\dot{a}{\,}-{\,}\dot{b}\bigr){\quad}.\cr}
\eqno(2.2)$$  
\noindent
The second term represents the non--linear coupling 
between a changing massless scalar field 
and a changing background geometry. 
A feature of this term is its resemblance to the action integral for 
$A^{(1)}(x)$, 
save for the 
${{1}\over{2}}({\dot{a}}-{\dot{b}})$ 
term in the integrand, 
which arises from the time--variation of the background metric. 
This interaction term is analogous to the phase and area/number 
interaction discussed in [9], 
which is there regarded as leading to the Hawking effect.  
The phase term is represented by the action,
while the 
${{1}\over{2}}({\dot{a}}-{\dot{b}})$ 
term is effectively the black--hole energy width, 
which describes the rate of change of the mass 
$m(t,r)$ 
inside a radius 
$r$ 
at time 
$t$ 
in the Vaidya--like region, 
where 
$e^{-a}{\;}{\,}
\simeq{\;}{\,}e^{b}{\;}{\,}
\simeq{\;}{\,}1{\,}-{\,}2Gm(t,r)/r{\,}$;
this term is identified with the area. 
Since we are here only concerned with physics 
as measured on the final hypersurface 
$\Sigma_{F}$ 
at late times 
$T$, 
the width term will be slowly varying and of order 
$O(m^{-3})$.  
If one removes it from the integrand 
and estimates the action term as being 
$O(m^{2})$ 
(on dimensional grounds), 
one finds that the interaction term is 
$O(m^{-1})$.  
In [9], 
the interaction term was indeed shown to be 
$O(m^{-1})$,
thus giving a small contribution for large 
$m$. 
Writing 
$\omega$ 
for a typical mode frequency of the perturbed scalar field [10], 
one finds that, 
in the adiabatic approximation
$\omega{\,}\gg{\,}
{{1}\over{2}}{\,}\bigl\vert{\dot{a}}-{\dot{b}}\bigl\vert{\,}$, 
the energy 
$E$ 
decomposes approximately into a sum over modes.

Including now also spin--1 and spin--2 particles, 
with their polarisation states, 
the total energy of the final radiation field, 
in the adiabatic approximation, 
is
$$E{\quad}
={\quad}M{\,}c^{2}{\quad}
={\quad}\hbar{\,}\sum_{s=0,1,2}{\;}{\,}\sum_{\ell=s}^{\infty}{\;}{\,}
\sum_{P}{\;}{\,}\sum_{m=-\ell}^{\ell}{\;}{\,}
{{(\ell -s)!}\over{(\ell +s)!}}{\;}{\,}
\int_{0}^{\infty}d\omega{\;}{\;}\omega{\;}{\;}
{\bigl\vert}b_{s\omega\ell mP}{\bigl\vert}^{2}
{\quad},
\eqno(2.3)$$  
\noindent
where
$M$
is the initial 
$ADM$ 
mass.

The amplitudes 
$\{b_{s\omega\ell mP}\}$ 
relate to the relic radiation reaching 
$\cal{I}^{+}{\,}$, 
and can be expressed in terms of final amplitudes 
associated with the space--like hypersurface 
$\Sigma_{F}$ 
[10].
As described in [3], 
our boundary--value approach leads to a discrete set 
of final radiation frequencies, 
given in the adiabatic approximation by 
$\{{\,}\omega_{n}{\,}
={\,}2{\pi}{\,}n{\,}{\hat{T}}^{-1}{\;};
{\;}n=0,1,2,{\,}\ldots\}$, 
where 
$\hat{T}{\,}={\,}2{\,}\bigl\vert{T}\bigl\vert$ 
and where 
$T$ 
is the 
(complex) 
asymptotic time--interval at spatial infinity, 
between the initial and final boundary hypersurfaces
$\Sigma_{I}$ 
and 
$\Sigma_{F}{\,}$.  
This gives 
$$\eqalign{E
{\quad}
= 
&{\quad}
\hbar{\,}
\sum_{s=0,1,2}{\;}{\,}\sum_{\ell=s}^{\infty}{\;}{\,}\sum_{P}{\;}{\,}
\sum_{m=-\ell}^{\ell}{\;}{\,}{{(\ell -s)!}\over{(\ell +s)!}}{\;}{\;}
\sum_{n=1}^{n_{\rm max}}{\;}{\;}{{2\pi}\over{\hat{T}}}{\;}{\;}
\omega_{n}{\;}{\;}
{\bigl\vert}b_{sn\ell mP}{\bigl\vert}^{2}\cr
=&{\quad}
{{2{\pi}\hbar}\over{\hat{T}}}{\;}{\;}
\sum_{n=1}^{n_{\rm max}}{\;}{\,}n{\;}{\,}N_{n}
{\quad},
\cr}
\eqno(2.4)$$
\noindent
where
$$N_{n}
{\quad}
=
{\quad}
{{2\pi}\over{\hat{T}}}{\;}{\,}\sum_{s=0,1,2}{\;}{\,}
\sum_{\ell=s}^{\infty}{\;}{\,}\sum_{P}{\;}{\,}
\sum_{m=-\ell}^{\ell}{\;}{\,}
{{(\ell -s)!}\over{(\ell +s)!}}{\quad}
{\bigl\vert}b_{sn\ell mP}{\bigl\vert}^{2}
{\quad}
\geq
{\quad}
0 
\eqno(2.5)$$
\noindent
is the occupation number of the mode 
$n$, 
and we have introduced a frequency cut--off through 
$n_{\rm max}{\,}$.
Therefore, 
the total excitation energy of the gas of particles 
is given in terms of the total oscillation quantum number 
$N$ 
as
$$E
{\quad}
=
{\quad}
E_{N}
{\quad}
=
{\quad}
{{hN}\over {\hat{T}}}
{\quad},
\eqno(2.6)$$
\noindent
where
$$N
{\quad}
=
{\quad}
\sum_{n=1}^{n_{\rm max}}{\;}n{\;}N_{n}
{\quad}.
\eqno(2.7)$$
\indent 
Thus, 
several states have the same total oscillation number. 
This equation describes combinatorially the splitting of 
$N$ 
into a sum of the numbers 
$n{\,}={\,}1{\,},{\;}{\ldots},{\,}n_{\rm max}{\,}$, 
with each number 
$n$ 
appearing 
$N_{n}$ 
times in the sum. 
Incidentally, 
Eq.(2.6) also gives the mean one--dimensional density of states
$$\rho(E)
{\quad}
=
{\quad}
{{\hat{T}(E)}\over{h}}
{\quad}
=
{\quad}
{{V}\over{2\pi\hbar c}}
{\quad},
\eqno(2.8)$$
\noindent
where 
$\hat{T}$ 
is the period of the phase--space trajectory with fixed energy 
$E$
and
$V=c{\,}\hat{T}$,
defined below in Eq.(3.1).\par
\smallskip
\indent 
Naturally, 
the quantity 
$n_{\rm max}$ 
is given by noting that the radiation energy 
should not exceed the total energy 
$E$.  
That is,
$$\hbar{\;}\omega_{n_{\rm max}}
{\quad}
=
{\quad}
E
{\quad},
\eqno(2.9)$$
\noindent
whence 
$n_{\rm max}{\,}={\,}N$. 
Thus, 
Eq.(2.7) implies that the ensemble average value of
$N_{n}$ 
is unity:
$${{1}\over{N}}{\;}{\,}\sum_{n=1}^{N}{\;}{\,}n{\;}N_{n}
{\quad}
=
{\quad}
1
{\quad}.
\eqno(2.10)$$
\indent 
By ergodicity, 
the ensemble average is equal to the time average for large 
$N$. 
This result follows from Eqs.(2.6,10), 
on substituting 
$N$ 
for 
$\hat{T}$, 
and then taking the continuum limit in the summation.\par
\smallskip
\indent 
If the relic Hawking--radiation state 
is independent of the azimuthal quantum number 
$m$, 
then one can write
$$N_{n}
{\quad}
=
{\quad}
{{2\pi}\over{\hat{T}}}{\;}
\sum_{s=0,1,2}{\;}{\,}
\sum_{P}{\;}{\,}
\sum_{\ell=s}^{\infty}{\;}{\;}
c_{s\ell}{\;}{\;}
{\bigl\vert}b_{sn\ell P}{\bigl\vert}^{2}
{\quad},
\eqno(2.11)$$
\noindent
where
$$c_{s\ell}
{\quad}
=
{\quad}
\bigl(2{\ell}+1\bigr){\;}{\;}{{(\ell-s)!}\over{(\ell+s)!}}
{\quad},
\eqno(2.12)$$
\noindent
allowing for the degeneracy in the angular momentum modes. 
With a view to simplifying the calculation, 
a further assumption would be to approximate the black hole 
as a point radiator, 
which emits only s--waves. 
S--wave emission is only sensitive to the radial coordinate, 
and higher angular modes are less significant. 
In general, 
a black body of radius 
$R$ 
emits modes with angular momenta 
${\,}\ell{\;}\leq{\;}\ell_{\rm max}{\,}={\,}\omega{\,}R{\,}$, 
where 
$\omega$ 
is the radiation frequency. 
This would introduce a cut--off in the 
$\ell$ 
summation.\par
\bigskip
\noindent 
{\bf 3. Density of States}
\medskip
\indent 
As is well known, 
the canonical--ensemble approach to black--hole evaporation 
breaks down in Schwarzschild space--time, 
as the Hawking temperature, 
$T_{H}{\,}$, 
is inversely related to the black--hole mass, 
$M$,
whence the specific heat 
$C={\partial M}/{{\partial}T_{H}}$ 
is negative. 
Another way of seeing this is to note that the density 
of black--hole states increases roughly at the rate 
$e^{M^{2}}$, 
for large 
$M$,
resulting in a canonical partition function which diverges at large 
$M$. 
Therefore, 
one should instead employ the microcanonical ensemble,
which involves a fixed total energy 
$E{\,}={\,}M{\,}c^{2}{\,}$; 
in this case, 
the total energy is not permitted to fluctuate.  
The microcanonical entropy does, 
however, 
involve corrections to the Bekenstein--Hawking area law, 
due to quantum space--time fluctuations at fixed horizon area 
(equivalent to fixed energy), 
as seen, 
for example,
in the loop quantum gravity approach [11].
In Sec.5.2, 
we discuss the nature of these entropy corrections.

In this Section, 
we calculate the microcanonical total density of states 
$\Omega(E,V)$.  
For an isolated system, 
this counts the number of microscopic configurations in a volume
$$V
{\quad}
=
{\quad}
c{\;}\hat{T}
{\quad},
\eqno(3.1)$$
\noindent
possessing energies distributed between zero and 
$E$. 
The energy constraint is the only conservation law that we impose.
In the quantum theory, 
when one is counting, 
different degrees of freedom correspond to distinct states.\par
\smallskip
\indent 
Another way of interpreting the microcanonical density of states is as
$$\Omega(E,V)
{\quad}
=
{\quad}
\sum_{\acute{N}=1}^{\infty}{\;}{\,}
\Omega_{\acute{N}}(E,V)
{\quad},
\eqno(3.2)$$
\noindent
where 
$\Omega_{\acute{N}}(E,V)$,
as below, 
gives the number density of states 
$\{N_{\acute{N}}\}$
which contain 
${\acute{N}}$ 
particles.
{~}These 
${\;}{\acute{N}}$ 
{~}particles {~}may {~}be {~}labelled {~}by {~}frequencies,
{~}ordered {~}such {~}that
${\omega}_{1}{\;}\leq{\;}{\omega}_{2}{\;}\leq{\;}
\ldots{\;}\leq{\;}{\omega}_{\acute{N}}{\;}$.
In Eq.(3.2), 
${\,}\Omega(E,V){\,}$ 
gives the phase--space volume, 
up to the energy 
$E{\,}$, 
of the system with volume 
$V{\,}$, 
where 
$\acute{N}$ 
denotes the number of degrees of freedom. 
We are effectively considering a set of oscillators labelled by 
$\{b_{j}\}$,
whose phases are random; 
that is, 
their location in phase space is unknown 
or their phase--space trajectories are uncorrelated. 
If the phases were not random, 
the motion of the system would not be ergodic, 
and entropy could not be defined.\par
\smallskip
\indent 
In order to determine 
$\Omega(E,V)$, 
we first define
$$\lambda(E,V)
{\quad}
=
{\quad}\
sum_{\{N_{n}\}}{\;}{\,}
\theta\Bigl(E{\,}-{\,}\hbar{\,}\sum_{n}{\;}\omega_{n}{\,}N_{n}{\,}\Bigr)
{\quad},
\eqno(3.3)$$
\noindent
where 
$\theta$ 
is the Heaviside step function 
and one sums over all states of the fields,
labelled by 
$\{N_{\acute{N}}\}$. 
The density of states is defined to be
$$\eqalign{\Omega(E,V)
{\quad} 
=
&{\quad}
{{\partial\lambda}\over{\partial E}}{\;}{\;}{\delta}E\cr
=
&{\quad}
{\delta}E{\;}{\,}\sum_{\{N_{n}\}}{\;}{\delta}
\Bigl(E{\,}-{\,}\hbar{\,}\sum_{n}{\,}\omega_{n}{\;}N_{n}\Bigr)
{\quad},
\cr}
\eqno(3.4)$$
\noindent
namely, 
the total number of configurations in the energy range 
$[E{\,},{\,}E{\,}+{\,}{\delta}E]$. 
The factor 
${\delta}E$ 
is some spread in energy or tolerance{~~} 
--- {~}independent of 
$E$, 
so as to make 
$\Omega(E,V)$ 
dimensionless{~~} 
--- {~}that depends on the precision with which the system 
has been prepared. 
One apparently natural condition is that 
$E{\,}\gg{\,}{\delta}E$, 
and that, 
for any energy fluctuation, 
${\Delta}E{\,}\gg{\,}{\delta}E$. 
For discrete systems, 
however, 
${\delta}E$ 
has a minimum value,
although for continuous systems it can be arbitrarily small{~~} 
--- {~}see Sec.5.2.\par
\smallskip
\indent 
The number density 
$\Omega_{\acute{N}}(E,V)$ 
may similarly be represented in the form:
$$\Omega_{\acute{N}}(E,V)
{\quad}
=
{\quad}
{\delta}E{\,}\sum_{\{N_{N'}\}}\delta
\biggl(E{\,}-{\,}\hbar{\,}\sum_{n=1}^{\acute{N}}{\;}{\,}
\omega_{n}{\;}N_{n}\biggr)
{\quad}.
\eqno(3.5)$$
\noindent
This expression (3.5) may also, 
in the continuum limit,
be re--written as:
$$\Omega_{\acute{N}}(E,V)
{\quad}
=
{\quad}
{{\delta E}\over {\acute{N}{\,}!}}{\;}{\,}
\int_{0}^{E}{\;}{\,}\prod_{n=1}^{\acute{N}}{\;}{\,} 
dE_{n}{\;}{\;}\rho(E_{n}){\;}{\;}
{\delta}\biggl(E{\,}-{\,}\sum_{n=1}^{\acute{N}}{\;}E_{n}\biggr)
{\quad},
\eqno(3.6)$$
\noindent
where 
$E_{n}=\hbar{\,}\omega_{n}{\,}N_{n}{\,}$, 
and where 
$\rho(E_{n})$ 
(the single--particle degeneracy) 
is given by Eq.(2.8). 
The factor 
$\prod_{n=1}^{\acute{N}}{\,}dE_{n}{\,}\rho(E_{n})$ 
counts the number of 
$\acute{N}$--particle states 
such that the $n$--th particle has energy 
$E_{n}$. 
The 
$1/\bigl(\acute{N}{\,}!{\,}\bigr)$ 
term accounts for identical particles in classical statistics, 
while the delta--function constraint counts the number 
of $N$--particle states with energy 
$E{\,}$; 
for further details, 
see [12].\par
\smallskip
\indent 
Thus, 
we consider the set of states 
$\{N_{n}\}$ 
as comprising subsets, 
labelled by 
$\acute{N}$, 
such that 
$\Omega_{\acute{N}}(E,V)$ 
is the number density of states possessing 
$\acute{N}$ 
particles.  
One can then evaluate 
$\Omega_{\acute{N}}(E,V)$, 
giving
$$\eqalign{\Omega_{\acute{N}}(E,V)
{\quad} 
&=
{\quad}
{\delta}E{\;}{\;}{{E^{\acute{N}-1}}\over{(\acute{N}-1)!}}{\;}{\;}
\prod_{n=1}^{\acute{N}}{\;}{\,}
{{1}\over{\hbar{\,}\omega_{n}}}\cr
&=
{\quad}
\biggl({{\delta E}\over{E}}\biggr){\;}{\,}
{{1}\over{\acute{N}{\,}!{\,}(\acute{N}-1)!}}{\;}{\;}
\biggl({{\hat{T}{\,}E}\over{\hbar}}\biggr)^{\acute{N}}
{\quad}.
\cr}
\eqno(3.7)$$
\noindent
Now, 
define a 
(small) 
dimensionless parameter 
$g$ 
by:
$$g^{2}
{\quad}
=
{\quad}
{{2\pi{\,}G{\,}M}\over{c^{2}{\,}V}}
{\quad}
=
{\quad}
\biggl({{M}\over{m_{p}}}\biggr)^{2}{\;}{\,}N^{-1}
{\quad},
\eqno(3.8)$$
\noindent
where the second equality comes from using Eqs.(2.6, 3.1).
We then obtain the energy spectrum [13]:
$$E
{\quad}
=
{\quad}
g{\,}E_{p}{\,}\sqrt{N}
{\quad},
\eqno(3.9)$$
\noindent
where 
$E_{p}
=\bigl({\hbar}c^{5}/G\bigr)^{{{1}\over{2}}}$ 
is the Planck energy. 
Notice that, 
for 
$E{\,}\gg{\,}E_{p}{\,}$,
the density 
$\Omega_{\acute{N}}(E,V)$ 
increases as 
$g^{2}$ 
decreases, 
corresponding to the notion that the phase-space volume{~~} 
--- {~}and therefore also the entropy{~~} 
--- {~}increases as the observation time 
$\hat{T}$ 
exceeds the dynamical time--scale of the collapsing matter, 
as the black hole forms and loses its hair. 
In a related calculation in a cosmological context in [14], 
the aim was to find a gravitational entropy function 
which increases monotonically as the system 
becomes more inhomogeneous.\par
\smallskip
\indent 
The microcanonical entropy is then given by:
$$S(E,V)
{\quad}
=
{\quad}
k_{B}{\;}{\,}\log\Biggl(\sum_{\acute{N}=1}^{\infty}{\;}{\,}
\Omega_{\acute{N}}(E,V)\Biggr)
{\quad}.
\eqno(3.10)$$
\noindent
One can approximate the right--hand side of Eq.(3.10) as
$$S(E,V)
{\quad}
\sim
{\quad}
k_{B}{\;}{\,}\log\Bigl(\Omega_{\hat{N}}(E,V)\Bigr)
{\quad},
\eqno(3.11)$$
\noindent
where 
$\hat{N}$ 
is the value of 
$\acute{N}$ 
which maximises
$\log\Bigl(\Omega_{\acute{N}}(E,V)\Bigr)$; 
that is, 
the most probable value of 
$\acute{N}{\;}$:
$${{\partial\Omega_{\acute{N}}}\over{\partial\acute{N}}}
{\biggl\vert}_{\acute{N}}
{\quad}
=
{\quad}
\hat{N}(E,V)
{\quad}
=
{\quad}
0
{\quad}.
\eqno(3.12)$$
\noindent
Using Stirling's formula for large 
$\acute{N}$, 
namely,
$\acute{N}!
{\;}
\sim
{\;}
\sqrt{2\pi}{\;}e^{-\acute{N}}{\,}
\acute{N}^{\acute{N}+{{1}\over{2}}}{\,}$, 
and assuming that 
${\delta}E/E$ 
is independent of 
$\acute{N}$ 
(see Sec.5.2), 
we obtain
$$\hat{N}
{\quad}
\simeq
{\quad}
\sqrt{{EV}\over{{\hbar}c}}
{\quad} 
\gg
{\quad}
1
{\quad}.
\eqno(3.13)$$
\noindent
Note that 
$\hat{N}=1$ 
corresponds to an elementary particle, 
as 
$\hat{T}$ 
is then the black--hole Compton time.
Eq.(3.13) seems to be related to the proposal in [15] 
for the maximum number of space--time bits 
in a spatial region with total volume 
$V$ 
and energy 
$E$. 
Therefore, 
the degeneracy is given by
$$\Omega(E,V)
{\quad}
\sim
{\quad}
\exp\Biggl(2{\,}\sqrt{{EV}\over{{\hbar}c}}\Biggr)
{\quad},
\eqno(3.14)$$
\noindent
and
$$S(E,V)
{\quad}
\sim
{\quad}
2{\,}k_{B}{\;}\sqrt{{EV}\over{{\hbar}c}}
{\quad} 
=
{\quad}
2{\,}k_{B}{\,}\hat{N}
{\quad}
=
{\quad}
2{\,}k_{B}{\,}\sqrt{N}
{\quad}.
\eqno(3.15)$$
\noindent
This is one of the main results of this paper. 
Note that the entropy is independent of the gravitational constant 
$G{\,}$.\par
\smallskip
\indent 
In arriving at Eq.(3.15), 
the 
${\delta}E/E$ 
term was neglected in comparison with the large 
$\hat{N}$ 
term. 
This will be justified in Sec.5.2. 
Thus, 
we find the well--known result that, 
for ideal bosons in a one--dimensional harmonic--oscillator potential, 
the asymptotic density of states,
as 
$N\to\infty{\,}$, 
is equal to the number of ways of partitioning an integer 
$N$ 
into a sum of other integers [16].\par
\smallskip
\indent
Through a little manipulation, 
the entropy may be expressed in the form
$$S
{\quad}
=
{\quad}
{{2{\,}k_{B}}\over{g}}{\;}
\biggl({{E}\over{E_{p}}}\biggr)
{\quad}.
\eqno(3.16)$$
\noindent
Thus, 
the entropy is linear in the total energy 
$E{\,}$. 
The 
$g$--dependence illustrates that, 
as 
$g\to{0}$, 
the entropy increases, 
as one expects in the classical limit.\par
\bigskip
\noindent
{\bf 4. Black--Hole/String Correspondence}
\medskip
\indent
One could compare the results in Eqs.(3.9,15) with those in [17],
which assert that the entropy of a Schwarzschild black hole 
in string theory is proportional to 
$\sqrt{N_{s}}{\,}$,
where 
$N_{s}$ 
is the level number of a long excited string which collapses 
into a Schwarzschild black hole at a critical string coupling.

Susskind [18] originally developed the idea 
that there is a one--to--one correspondence 
between quantum Schwarzschild black holes 
and fundamental string states, 
building upon the ideas first developed in [19]. 
One immediate similarity is that the degeneracy of states 
for a given mass increases very rapidly with mass, 
both for black holes and for fundamental strings. 
Yet, 
the leading string entropy term depends linearly upon the mass, 
whereas the Bekenstein--Hawking entropy is quadratic. 
The Correspondence Principle [17], 
however, 
aims to match these differing behaviours 
for a particular string coupling 
$g_{c}{\,}$. 
A related 
(semi--classical) 
quantum field theory/quantum string black hole duality 
was investigated in [20].\par
\smallskip
\indent 
The one--dimensional string is characterised by its tension
$\sim{\,}(\ell_{s})^{-2}$, 
where 
$\ell_{s}$ 
is the string scale, 
and by a 
(dimensionless) 
interaction strength 
$g{\,}$. 
Varying the coupling 
$g$ 
has an interpretation in terms of a varying dilaton background, 
where the dilaton is given by a scalar field 
$\phi{\,}$, 
through the relation 
$g^{2}=e^{\phi}$. 
In practice, 
we treat 
$\phi$ 
as a constant and 
$g$ 
as a parameter. 
Consider a highly--excited string state and a weak string coupling 
$g{\,}\ll{\,}1{\,}$, 
where, 
in four dimensions, 
$\ell_{s}$ 
and the Planck scale 
$\ell_{p}{\,}={\,}\sqrt{G{\hbar}/c^{3}}$ 
are related through
$$g
{\quad}
=
{\quad}
{{\ell_{p}}\over{\ell_{s}}}
{\quad}.
\eqno(4.1)$$
\indent
Thus, 
we initially consider a string scale far above the Planck scale. 
By increasing the coupling adiabatically, 
(in effect, 
then, 
by increasing the strength of the gravitational interaction), 
one reaches a point such that the Schwarzschild radius of the string,
given in terms of the string mass scale 
$M_{s}$ 
by 
$2GM_{s}/c^{2}{\,}$,  
becomes larger than 
$\ell_{s}{\,}$; 
at this point, 
the string makes a 
(smooth) 
transition into a black hole. 
Conversely, 
however, 
had one started with a black hole, 
and then decreased the coupling, 
one would find that the size of the black hole 
would eventually fall below the string scale. 
The correspondence point would be that at which the black hole 
makes the transition to a string.\par 
\smallskip
\indent
The 
(closed) 
string mass spectrum has the form [17]
$$M
{\quad}
\sim
{\quad}
{{\hbar{\,}\sqrt{N_{s}}}\over{c{\,}\ell_{s}}}
{\quad}
=
{\quad}
g{\,}m_{p}{\,}\sqrt{N_{s}}
{\quad},
\eqno(4.2)$$
\noindent
where 
$m_{p}=\hbar/\bigl(\ell_{p}c\bigr)$ 
is the Planck mass, 
and 
$N_{s}{\,}\gg{\,}1$
denotes the level number of a long 
(highly--excited) 
string. 
Therefore, 
for fixed 
$N_{s}{\,}$, 
the string mass increases as 
$N_{s}$ 
increases. 
The string spectrum involves an infinite tower of massive states, 
as well as gravitons and massless spin--one particles 
(for example). 
For states in free string theory, 
with a given mass, 
the most numerous states are the single--string states, 
since the string entropy depends linearly upon energy. 
The resemblance of Eq.(4.2) to Eq.(3.9) is evident.\par 
\smallskip
\indent
The Bekenstein--Hawking entropy 
{~}of {~}a {~}four--dimensional {~}Schwarzschild {~}black {~}hole, 
$S_{B-H}{\,}$,  
is given in terms of 
$\ell_{p}$ 
as
$$S_{B--H}
{\quad}
=
{\quad}
{{4{\pi}{\,}k_{B}{\,}M^{2}{\,}\bigl(\ell_{p}\bigr)^{2}{\,}c^{2}}
\over{\hbar^{2}}}
{\quad},
\eqno(4.3)$$ 
\noindent 
where 
$M$ 
is the black--hole mass. 
Thus, 
at the correspondence point where 
$M=M_{s}$ 
(we assume that the mass does not change over the transition), 
Eq.(4.3) becomes
$$S_{B--H}
{\quad}
\sim
{\quad}
{{4{\pi}{\,}k_{B}{\,}\bigl(\ell_{p}\bigr)^{2}{\,}N_{s}}
\over{\bigl(\ell_{s}\bigr)^{2}}}
{\quad}
=
{\quad}
4{\pi}{\,}k_{B}{\,}g^{2}{\,}N_{s}
{\quad}.
\eqno(4.4)$$ 
\noindent
Therefore, 
at the critical coupling [17],
$$\bigl(g_{c}\bigr)^{2}
{\quad}
\sim
{\quad}
\bigl(N_{s}\bigr)^{-{{1}\over{2}}}
{\quad},
\eqno(4.5)$$
\noindent
the Bekenstein--Hawking entropy becomes the string entropy
$$S_{\rm string}
{\quad}
\sim
{\quad}
4{\pi}{\,}k_{B}{\,}\sqrt{N_{s}}
{\quad},
\eqno(4.6)$$
\noindent
which resembles Eq.(3.15). 
One sees that the degrees of freedom of an excited free string 
can match the entropy of a four--dimensional Schwarzschild black hole. 
It can also be demonstrated that the sub--leading entropy terms, 
which depend upon the mass, 
also agree [21].  
In this case, 
the degrees of freedom of a perturbatively interacting string 
can be identified with Schwarzschild black--hole states.\par 
\smallskip
\indent
Instead {~}of {~}fixing 
${\;}g{\;}$, 
{~}and {~}letting {~}the {~}black--hole {~}mass 
${\;}M$ 
{~}(and {~}therefore 
${\,}N_{s}{\;}$) 
{~}vary {~}as {~}the {~}black {~}hole {~}evaporates, 
{~}one {~}can, 
{~}equivalently, 
{~}keep {~}the {~}entropy {~}fixed, 
{~}such {~}that 
${\,}S_{\rm string}{\,}={\,}S_{B-H}$ 
{~}(equivalent {~}to {~}keeping {~}the {~}state {~}fixed), 
{~}while {~}varying 
$g$ 
{~}adiabatically {~}[22].  
{~}Note {~}that, 
{~}for 
${\;}{\,}g^{2}{\,}<{\,}\bigl(N_{s}\bigr)^{-{{1}\over{2}}}{\,}$,
{~}one {~}has
${\;}{\,}S_{\rm string}{\,}>{\,}S_{B-H}{\;}$, 
{~}and {~}conversely, 
{~}for {~}the {~}case 
${\;}{\,}g^{2}{\,}>{\,}\bigl(N_{s}\bigr)^{-{{1}\over{2}}}{\,}$, 
{~}one {~}has
${\;}{\,}S_{\rm string}{\,}<{\,}S_{B-H}{\;}{\,}$. 
{~}Thus, 
{~}if {~}there {~}is {~}some {~}critical {~}mass, 
${\;}M_{c}{\;}{\,}$, 
{~}such {~}that 
${\;}{\,}S_{\rm string}{\,}>{\,}S_{B-H}{\;}{\,}$, 
{~}then {~}a {~}black {~}hole {~}can {~}make {~}a {~}transition 
{~}to {~}a {~}higher--entropy {~}string {~}state. 
{~}This {~}mass 
${\;}M_{c}{\,}$ 
{~}satisfies 
${\;}{\,}GM_{c}/c^{2}{\;}\sim{\;}\ell_{s}{\;}{\,}$, 
{~}that {~}is,
$$M_{c}
{\quad}
=
{\quad}
M_{s}{\,}g^{-2}
{\quad}.
\eqno(4.7)$$
\noindent
This mass scale is much larger than the Planck mass 
$m_{p}{\,}$, 
in the case that 
$g{\,}\ll{\,}1$.
Further, 
at this mass, 
the Hawking temperature equals the Hagedorn temperature
$$T_{s}
{\quad}
=
{\quad}
\bigl({\hbar}c\bigr)/\bigl(k_{B}{\,}b{\,}\ell_{s}\bigr)
{\quad},
\eqno(4.8)$$
\noindent
where 
$b$ 
is a constant which depends on the particular model 
and on the dimension taken for space--time [18,19].  
Dimopoulos and Emparan [23] found that a black hole 
makes a transition to a state of highly--excited and jagged strings 
(a string ball) 
at the mass 
$M_{c}{\,}$.
The production cross sections of black holes 
and string balls can be matched at 
$M_{c}{\,}$. 
The string ball then loses mass through evaporation, 
and at a mass just below 
$M_{c}{\,}$, 
it `puffs up' to a larger random--walk size
$\sim{\,}\ell_{s}{\,}\bigl(M/M_{s}\bigr)^{{{1}\over{2}}}$ 
(see below [23]). 
The string ball continues to evaporate with a quasi--thermal spectrum, 
at the Hagedorn temperature, 
during which, 
its size reduces to 
$\sim{\,}\ell_{s}{\,}$. 
The black hole can be approximated as a point radiator, 
as its wavelength is of order 
$\bigl({\hbar}c\bigr)/\bigl(k_{B}{\,}T_{s}\bigr)$, 
which is larger than the size of the black hole. 
Since the specific heat of massive string modes is negative, 
these cannot remain in equilibrium with an infinite heat bath. 
Therefore, 
they evaporate into massless modes, 
so avoiding the possibility of a naked singularity [19].

In [19], 
the assumption 
was made that there is an intermediate heavy--string state, 
which decays at the Hagedorn temperature into massless strings; 
that is, 
into elementary particles such as photons. 
Zalewski [24], 
however, 
suggested that the black hole 
transforms directly into massless string states,
using statistical arguments 
and the property that the evanescent black hole 
should have zero angular momentum.\par 

In fact, 
one can also match the total width for string decay  
(into the dilaton, graviton and massless fields)
with that for quantum black holes [25]. 
The total string width has the form [26]:
$$\Gamma_{s}
{\quad}
=
{\quad}
{\lambda}{\,}g^{2}{\,}M
{\quad},
\eqno(4.9)$$
\noindent
where 
$\lambda$ 
is a dimensionless numerical factor. 
A formula analogous to Eq.(4.9) encapsulates the decay widths 
of all the Standard--Model particles, 
as well as of topological and non--topological solitons, 
and of cosmic strings [25]. 
The black--hole decay rate has the form
$$\Gamma_{BH}
{\quad}
=
{\quad}
{\biggl\vert}{{1}\over{M}}{\,}{{dM}\over{dt}}{\biggr\vert}
{\quad}.
\eqno(4.10)$$
\noindent
For the low--energy canonical ensemble, 
we use the Stefan--Boltzmann law 
$${{dM}\over{dt}}
{\quad}
\sim
{\quad}
-{\;}{{4{\pi}{\,}{\sigma}{\,}\bigl(R_{s}\bigr)^{2}{\;}
\bigl(k_{B}{\,}T_{BH}\bigr)^{4}}
\over{c^{4}{\,}{\hbar}^{2}}}
{\quad},
\eqno(4.11)$$
\noindent
where 
${\;}\sigma{\,}$ 
{~}is {~}a {~}dimensionless {~}constant, 
{~}the {~}Schwarzschild {~~}radius {~~}is {~~}denoted {~~}by
$R_{s}
=2GM/c^{2}{\,}$,
and 
$T_{BH}
=\bigl({\hbar}c^{3}\bigr)/\bigl(8{\pi}GMk_{B}\bigr)$ 
denotes the Hawking temperature. 
Thus,
$$\Gamma_{BH}
{\quad}
=
{\quad}
{{2{\,}\sigma{\,}\bigl(\ell_{p}\bigr)^{2}}
\over{{\hbar}^{2}{\,}c^{4}}}{\;}{\,}
\bigl(k_{B}{\,}T_{BH}\bigr)^{3}
{\quad}.
\eqno(4.12)$$
\noindent
As 
$T_{BH}{\,}\to{\,}T_{s}$ 
and 
$R_{s}{\,}\to{\,}\ell_{s}{\,}$; 
that is, 
as the black hole's temperature 
increases towards the string temperature during evaporation, 
then
$$\Gamma_{BH}
{\quad}
\to
{\quad}
{{2\sigma}\over{b^{3}}}{\;}g^{2}{\;}M_{s}
{\quad},
\eqno(4.13)$$
\noindent
which resembles Eq.(4.9) in the case that 
$M{\,}\sim{\,}M_{s}{\,}$.
\par 
\smallskip
\indent 
A random--walk interpretation of strings 
at zero coupling also exists [22,27].
From Eq.(4.6), 
with 
$n=\sqrt{N_{s}}{\,}$,
the string entropy can be represented as an integer 
$n$, 
up to a factor, 
giving the number of random--walk steps or number of string bits 
in a polymer string representation [28]. 
Each step has length 
$\ell_{s}$
and mass 
$M_{s}{\,}$. 
At zero coupling, 
the free string effectively has a random--walk size 
$\sqrt{n}{\;}\ell_{s}
=(N_{s})^{{{1}\over{4}}}{\,}\ell_{s}
{\,}
\gg
{\,}
\ell_{s}{\,}$, 
and the total length of the string is 
$L=n\ell_{s}{\,}$.
\par
\bigskip
\noindent 
{\bf 5. Thermodynamics}
\medskip
\indent 
Evidently, 
from Eq.(3.15), 
$S(E,V)$ 
is an extensive function of 
$E$ 
and 
$V$, 
both of which determine the global properties 
of the system at infinity. 
The corresponding thermodynamical quantities, 
such as temperature 
$\theta$, 
pressure 
$P$, 
specific heat 
$C_{V}$ 
and average total particle number 
$\bar{N}$, 
can be computed using the standard formulae
$$\left({{\partial S}\over{\partial E}}\right)_{V}
\equiv{\;}{{1}\over{\theta}}{\;},
{\quad}\beta{\,}P{\;}
\equiv{\;}\left({{\partial S}\over{\partial V}}\right)_{E}{\;},
{\quad}C_{V}{\;}
={\;}-\left[\theta^{2}
\left({{\partial^{2}S}\over{\partial
E^{2}}}\right)_{V}\right]^{-1}{\;},
{\quad}\bar{N}{\;}
={\;}{{{\beta}PV}\over{k_{B}}}
{\quad},
\eqno(5.1)$$
\noindent
where 
$\beta{\,}={\,}1/\bigl(k_{B}\theta\bigr)$, 
giving
$$\theta{\;}
={\;}{{1}\over{k_{B}}}{\,}\sqrt{{{E{\hbar}c}\over{V}}}{\;}
={\;}{{{\hbar}c}\over{k_{B}V}}{\,}\sqrt{N}{\;},
{\;}{\;}P{\;}
={\;}{{k_{B}E}\over{V}}{\;}
={\;}k_{B}{\,}\rho{\;},
{\;}{\;}C_{V}{\;}
={\;}S(E,V){\,}>{\,}0{\;},
{\;}{\;}\bar{N}{\;}
={\;}\hat{N}
{\,},
\eqno(5.2)$$
\noindent
where 
$\rho$ 
is the energy density. 
The second equation in Eq.(5.1) describes an incompressible fluid,
for which sound propagates at the speed of light{~~} 
--- {~}the Zel'dovich equation of state. 
We consider both 
$E$ 
and 
$V$ 
to be large, 
such that 
$\rho$ 
is finite and 
$E{\,}\gg{\,}{\hbar}c/V$.
The thermodynamic limit, 
in which 
$V$ 
is regarded as being arbitrarily large, 
should not be taken, 
as our approach is based on considering only finite time--intervals 
at infinity. 
Hence, 
the ideal gas approximation holds,
in that interactions between particles are being neglected: 
$$P{\,}V
{\quad}
=
{\quad}
\hat{N}{\,}k_{B}{\,}\theta
{\quad}.
\eqno(5.3)$$
\noindent
The number 
$\hat{N}
=E/\bigl(k_{B}{\theta}\bigr)$ 
is also the average number of particles with energy 
$k_{B}\theta$
in which the black hole of total energy 
$E$ 
decays. 
The property that 
$C_{V}{\,}>{\,}0$ 
implies that a stable thermal equilibrium 
can be reached by the radiation at any temperature.\par
\smallskip
\indent 
One can express the entropy in terms of the temperature 
$\theta$ 
and the volume or length 
$V$, 
as 
(see [29]):
$$S
{\quad}
=
{\quad}
{{(k_{B})^{2}}\over{\pi\hbar c}}{\;}{\,}\theta{\,}V
{\quad}.
\eqno(5.4)$$
\noindent
Thus, 
the entropy is proportional to the length 
$V$, 
and one may interpret the entropy as comprising bits of length 
$V$. 
One obtains the familiar expression for the entropy 
if one makes the replacement
$$V
{\quad}
\to
{\quad}
{{1}\over{6}}{\;}{\pi}^{2}{\,}f{\,}L
{\quad},
\eqno(5.5)$$
\noindent
where 
$L$ 
is a length scale and 
$f$ 
denotes the number of field species. 
Thus,
$$S
{\quad}
=
{\quad}
2\pi{\,}k_{B}{\;}\sqrt{{{fp}\over{6}}}
{\quad},
\eqno(5.6)$$
\noindent
where 
$p
=\bigl(EL\bigr)/\bigl(hc\bigr)$ 
is dimensionless. 
This is the Cardy formula for central charge 
$f$ 
[16].
Eq.(5.6) would have been obtained if, 
instead of summing over 
$\{N_{n}\}$ 
configurations, 
we had summed over 
$\{N_{n}^{i}\}$ 
configurations such that
$N_{n}
=\sum_{i=1}^{f}{\,}N_{n}^{i}{\,}$.
\par
\smallskip 
\indent 
When the particle energy 
$\hbar{\,}\omega_{n}$ 
is much larger than the corresponding temperature 
$\theta$, 
then, 
for 
$E{\,}\gg{\,}\hbar{\,}\omega_{n}{\,}$, 
one has
$$N
{\quad}
\gg
{\quad}
n
{\quad}
\gg
{\quad}
\sqrt{N}
{\quad},
\eqno(5.7)$$
\noindent
and the microcanonical ensemble should be used. 
Further, 
in the very--high--energy tail of the spectrum, 
where quanta have energies comparable with the total energy 
$E$ 
of the system, 
one has
$$n
{\quad}
\simeq
{\quad}
N
{\quad}.
\eqno(5.8)$$
\noindent
On the other hand, 
for low energies which obey the condition
$$n
{\quad}
\ll
{\quad}
\sqrt{N}
{\quad}
<
{\quad}
N
{\quad},
\eqno(5.9)$$
\noindent
the canonical ensemble description of the gas is valid [30].\par
\smallskip
\indent  
Another interesting relationship, 
following from Eq.(5.2), 
is
$${{V}\over{2\pi}}
{\quad}
=
{\quad}
{\hbar}{\,}c{\,}\beta{\,}\sqrt{N}
{\quad}
=
{\quad}
g^{-1}{\,}\ell_{p}{\;}\sqrt{N}
{\quad},
\eqno(5.10)$$
\noindent
which relates the wave period to the inverse temperature. 
Thus, 
in this representation, 
one interprets 
$V${~~} 
--- {~}effectively the time--separation at infinity{~~} 
--- {~}as a root--mean--square distance, 
in which case it represents the distance 
covered by a Brownian random walk with step--size
$g^{-1}{\,}\ell_{p}{\,}$, 
taking 
$N$ 
to be the total number of steps [31].  
The analogue of the string length, 
$\ell_{s}{\,}$, 
in our theory, 
is provided by 
$\hat{\ell}_{s}{\,}$, 
where
$$\hat{\ell}_{s}
{\quad}
=
{\quad}
{{V}\over{2\pi}}{\;}N^{-{{1}\over{2}}}
{\quad}
=
{\quad}
{\hbar}{\,}c{\,}\beta
{\quad},
\eqno(5.11)$$
\noindent
which is just the random--walk step size. 
In addition, 
with
$\hat{\ell}_{s}
={\hbar}/\bigl(\hat{M}_{s}{\,}c\bigr)$, 
the mass scale
$\hat{M}_{s}{\,}$, 
analogous to the string mass scale 
$M_{s}{\,}$, 
is
$$\hat{M}_{s}
{\quad}
=
{\quad}
{{M}\over{\sqrt{N}}}
{\quad}
=
{\quad}
g{\,}m_{p}
{\quad}.
\eqno(5.12)$$
\noindent
From Eq.(5.2), 
one has
$$\theta
{\quad}
=
{\quad}
{{{\hbar}c}\over{k_{B}{\,}\hat{\ell}_{s}}}
{\quad},
\eqno(5.13)$$
\noindent
which is the analogue of the Hagedorn temperature 
given in Eq.(4.8).\par
\smallskip
\indent 
In the light of Eq.(5.11), 
for the Euclidean quantum gravity approach to black--hole evaporation, 
the imaginary--time period is given by the inverse temperature 
of fields on the background space--time [32].\par
\bigskip
\noindent 
{\bf 5.1 Bekenstein-Hawking Entropy}
\medskip
\indent 
Given a four--dimensional, 
non--rotating, 
neutral black hole with horizon area 
$A$, 
the Bekenstein--Hawking entropy has the well--known form
$$S_{B--H}
{\quad}
=
{\quad}
{{k_{B}A}\over{4\bigl(\ell_{p}\bigr)^{2}}}
{\quad}
=
{\quad}
4{\pi}{\,}k_{B}\left({{M}\over{m_{p}}}\right)^{2}
\eqno(5.14)$$
\noindent
This is a semi--classical result, 
valid for 
$A{\,}\gg{\,}\bigl(\ell_{p}\bigr)^{2}$. 
Let us now equate Eq.(3.15) with Eq.(5.14), 
which is valid for 
$M{\,}\gg{\,}m_{p}{\;}$:
$$S(E,V)
{\quad}
=
{\quad}
S_{B--H}
{\quad}.
\eqno(5.15)$$
\noindent
Matching the four--dimensional entropy with a two--dimensional entropy 
is just one way of comparing the thermodynamics 
of any two physical systems. 
Strong matching might involve equating partition functions which,
then, 
leads to the matching of all thermodynamical quantities [33]. 
The weak matching condition, 
Eq.(5.15), 
which does not necessarily imply the matching 
of other thermodynamical quantities,
implies that
$$\hat{T}(M)
{\quad}
=
{\quad}
\bigl(2\pi\bigr)^{3}{\;}{{G^{2}M^{3}}\over{{\hbar}{\,}c^{4}}}
{\quad}.
\eqno(5.16)$$
\noindent
In the semi--classical approximation, 
the evaporation time--scale,
$t_{0}{\,}$, 
of a non--rotating, 
neutral black hole of initial mass 
$M$, 
is [34]
$$t_{0}
{\quad}
=
{\quad}
\bigl(3\gamma\bigr)^{-1}{\;}{{G^{2}M^{3}}\over{{\hbar}{\,}c^{4}}}
{\quad},
\eqno(5.17)$$
\noindent
where 
$\gamma$ 
is a dimensionless coefficient, 
which depends on the number of particle species 
which can be emitted at a given temperature, 
and on grey--body factors.
Thus,
$\hat{T}$
is of the order of
$\hat{t_{0}}{\,}$.
\par
\smallskip
\indent
Another interesting way to arrive at a time--separation 
at infinity of the order of 
$t_{0}{\,}$, 
comes through setting, 
in Eq.(5.10):
$$g^{2}
{\quad}
=
{\quad}
\bigl(g_{c}\bigr)^{2}
{\quad}
=
{\quad}
{{1}\over{2{\pi}\sqrt{N}}}
{\quad}
\ll
{\quad}
1
{\quad},
\eqno(5.18)$$
\noindent
in which case we arrive back at Eq.(5.16). 
In other words,
the B--H entropy equals the 1--dimensional entropy Eq.(3.15), 
when Eq.(5.18) is satisfied. 
Therefore, 
we seem to have enough configurations on 
$\Sigma_{F}$ 
to reproduce the entropy of black holes. 
However, 
if one considered the case in which 
$2{\pi}{\,}g^{2}{\,}\sqrt{N}{\,}\gg{\,}1{\,}$, 
then one would still be in the black--hole phase, 
since 
$\hat{T}{\,}<{\,}t_{0}{\,}$.
When 
$2{\pi}{\,}g^{2}{\,}\sqrt{N}{\,}<{\,}1{\,}$, 
one is in a post--evaporation phase 
(perhaps string--like). 
Eq.(5.18) is essentially the critical coupling condition 
of Eq.(4.5).\par
\smallskip
\indent 
The microcanonical entropy 
in Eq.(3.15) exceeds the Bekenstein--Hawking value, 
if 
$\hat{T}$ 
exceeds the right--hand side of Eq.(5.16):
$$V
{\quad}
>
{\quad}
\left({{2{\pi}GM}\over{c^{2}}}\right)^{3}{\;}
\bigl(\ell_{p}\bigr)^{-2}
{\quad}.
\eqno(5.19)$$
\noindent
However, 
for 
$\hat{T}$ 
given by Eq.(5.16), 
one has, 
from Eq.(5.2),
$\theta=4T_{H}{\,}$, 
where 
$T_{H}$ 
is the Hawking temperature.\par
\smallskip
\indent 
We conclude that, 
if our final space--like hypersurface, 
$\Sigma_{F}{\,}$,
were located so that the time separation at spatial infinity 
between the initial and final hypersurfaces were given by 
${{1}\over{2}}\hat{T}$ 
in Eq.(5.16),
then the thermal entropy of the remnant radiation 
would be of the order of the initial black--hole entropy. 
In that case, 
therefore, 
the thermodynamics of the four--dimensional black hole 
would be encoded in that of a one--dimensional gas 
of non--interacting bosonic particles. 
There have been similar approaches that modelled black holes 
in any number of dimensions, 
using a one--dimensional gas of massless particles [35].\par
\smallskip
\indent 
As an aside, 
we mention here briefly the connection between our approach 
and the idea of strings 
growing or spreading upon reaching the black--hole stretched horizon,
which is located at a distance 
$\sim{\,}\ell_{s}$ 
from the event horizon [36]. 
One can re--write Eq.(3.8) in the following suggestive way:
$${{2{\pi}GM}\over{c^{3}}}
{\quad}
=
{\quad}
g^{2}{\,}\hat{T}
{\quad}.
\eqno(5.20)$$
\noindent
Setting 
$t_{\rm cross}
=2r_{s}/c{\,}$, 
which is the time--scale for light 
crossing a Schwarzschild black hole of radius 
$r_{s}=2GM/c^{2}{\,}$, 
then
$$t_{\rm cross}
{\quad}
=
{\quad}
g^{2}{\;}\biggl({{2{\,}\hat{T}}\over{\pi}}\biggr)
{\quad}.
\eqno(5.21)$$
\noindent
Due to the redshift effect, 
the string and all the information it carries appears, 
from the perspective of a distant observer, 
to grow as the string passes through the stretched horizon, 
while, 
to a freely--infalling observer, 
it appears to be a Planck--sized object 
as it crosses the stretched horizon. 
The string will spread across the entire black hole in a time [36]
$$t_{\rm spread}
{\quad}
=
{\quad}
\left({{\ell_{p}}\over{\ell_{s}}}\right)^{2}{\;}
{{G^{2}M^{3}}\over{{\hbar}{\,}c^{4}}}
{\quad}
=
{\quad}
g^{2}{\;}{{G^{2}M^{3}}\over{{\hbar}{\,}c^{4}}}
{\quad},
\eqno(5.22)$$
\noindent
where 
$g$ 
is here defined as in Eq.(3.17).
The time 
$t_{\rm spread}$
is short, 
compared to the black--hole evaporation time 
$t_{0}{\,}$, 
provided that 
$g$ 
is small. 
One can immediately see the similarities between Eqs.(5.21) and (5.22),
in the case that 
$\hat{T}$ 
is of the order of the black--hole evaporation time. 
Furthermore, 
both 
$t_{\rm cross}$ 
and 
$t_{\rm spread}$
seem to encapsulate the notion of information 
traversing the entire black hole.\par
\bigskip
\noindent
{\bf 5.2 Corrections to the Microcanonical Entropy}
\medskip
\indent 
Corrections to the B--H area law are important, 
and related to the evaporation of the black hole. 
These corrections to the microcanonical entropy, 
$S_{MC}{\,}$, 
due to quantum fluctuations at fixed horizon area 
$A{\,}$, 
for 
$A{\,}\gg{\,}\bigl(\ell_{p}\bigr)^{2}$, 
seem to have a 
{\it universal} 
form, 
that is, 
a form which is independent of the particular parameters 
of the black hole [37]:
$$S_{MC}
{\quad}
=
{\quad}
S_{B-H}{\;}
-{\;}{\,}{{3}\over{2}}{\,}\log\Bigl(S_{B-H}\Bigr){\;}
+{\;}{\,}{\rm constant}{\;}
+{\;}{\,}O\Bigl(\bigl(S_{B-H}\bigr)^{-1}\Bigr)
{\quad},
\eqno(5.23)$$
\noindent
where 
$S_{B--H}$ 
is given in Eq.(5.14). 
Such corrections 
can also be interpreted as being due to thermal fluctuations 
about the black hole's equilibrium configuration. 
The 
${\;}-{\,}{{3}\over{2}}{\,}\log\bigl(S_{B-H}\bigr)$ 
term, 
in particular, 
holds also for all black holes whose microscopic degrees of freedom 
are described by an underlying conformal field theory, 
including BTZ and string--theoretic black holes [37,38].
This term also appears when one computes a corrected Cardy formula 
for the density of states [39]{~~} 
--- {~}see also [41].\par
\smallskip
\indent 
We now compute the corrections to Eq.(3.15), 
of the type given in Eq.(5.23), 
which arise from quantum fields on the black--hole background 
(as studied above),
and from the reduction in the area or mass of the black hole.
The B--H law Eq.(3.15) is, 
of course, 
only valid for 
$A{\,}\gg{\,}\bigl(\ell_{p}\bigr)^{2}$. 
One may either convert the 
$\acute{N}$ 
sum in Eq.(3.10) into an integral, 
and then perform a stationary--phase approximation, 
or one may compute the sum exactly. 
A straightforward calculation yields
$$S(E,V)
{\quad}
\sim
{\quad}
2k_{B}{\;}\sqrt{{{EV}\over{hc}}}{\;}{\,}
+{\;}{\,}{{1}\over{2}}{\;}k_{B}{\,}
\log\Biggl(\sqrt{{{EV}\over{hc}}}\Biggr){\;}{\,}
+{\;}{\,}k_{B}{\,}\log\biggl({{{\delta}E}\over{E}}\biggr){\;}{\,}
+{\;}{\,}{\rm constant}
{\quad},
\eqno(5.24)$$
\noindent
where we have neglected higher--order terms. 
The 
$+{\,}{{1}\over{2}}$ 
coefficient of the logarithm 
in Eq.(5.24) is in contrast to the results in [42]. 
It has been suggested that different quantum theories 
lead to different quantum corrections 
to the black--hole entropy [37].\par
\smallskip
\indent 
One can, 
however, 
make further progress with the 
$\log\bigl({\delta}E/E\bigr)$ 
term in Eq.(5.24). 
Since the energy levels are discrete, 
${\delta}E$ 
has a minimum value, 
which one may take, 
naturally, 
to be the spacing between the particle energy levels:
$$\hbar{\,}\bigl({\,}{\omega}_{n+1}{\,}-{\,}{\omega}_{n}\bigr)
{\quad}
=
{\quad}
{h\over{\hat{T}}}
{\quad}.
\eqno(5.25)$$
\noindent
Thus, 
for 
$V{\,}\gg{\,}\bigl(hc\bigr)/\bigl(4E\bigr)$ 
and for
$${\delta}E
{\quad}
=
{\quad}
\hbar{\,}{\omega}_{0}
{\quad}
=
{\quad}
{{hc}\over{V}}
{\quad},
\eqno(5.26)$$
\noindent
(that is, 
the condition that the energy spread depends solely on the volume),
we find, 
up to unimportant constant terms, 
that
$$\Omega(E,V)
{\quad}
=
{\quad}
2^{-{{3}\over{2}}}{\,}
\left({{EV}\over{hc}}\right)^{-{{3}\over{4}}}
\exp\left(2{\,}\sqrt{{EV}\over{hc}}\right)
{\quad},
\eqno(5.27)$$
\noindent
which implies that
$$S(E,V)
{\quad}
\sim
{\quad}
k_{B}
\left(2{\,}\sqrt{N}{\,}-{\;}{{3}\over{2}}{\,}
\log\Bigl(2{\,}\sqrt{N}\Bigr)\right){\,}
+{\;}{\,}{\rm constant}
{\quad}.
\eqno(5.28)$$
\noindent
In terms of the leading 
${\;}-{\,}{{3}\over{2}}{\,}\log({~~~})$ 
term, 
this agrees with [37,39], 
provided that we identify Eq.(3.15) with the black--hole entropy. 
Clearly, 
the discreteness of the energy levels is a crucial element 
in deriving Eq.(5.28).\par
\smallskip
\indent 
One might compare Eq.(5.28) with the calculation of the logarithmic 
corrections to the Cardy formula for the density of states 
in a two--dimensional conformal field theory [39].  
In [39], 
there is an arbitrary period, 
which is chosen so that the central charge is 
{\it universal} 
{~}(independent {~}of {~}black--hole {~}area); 
{~}this {~}again {~}leads {~}to {~}the 
${\;}-{\,}{{3}\over{2}}{\,}\log\bigl(S_{B--H}\bigr)$ 
term{~~}
--- {~}see also [40].
A similar conclusion holds in the formulation of the present paper,
where the spread in energy is fixed in relation to the period 
at infinity between the initial and final space--like hypersurfaces, 
$\Sigma_{I}$ 
and 
$\Sigma_{F}{\,}$, 
as in Eq.(5.26).\par
\bigskip
\noindent 
{\bf 6. Probability Distribution}
\medskip
\indent 
We now discuss further our previous results 
for quantum amplitudes in black--hole evaporation [3]. 
We set approximately spherically--symmetric boundary data 
on an initial space--like hypersurface, 
$\Sigma_{I}{\,}$, 
and weak, 
non--spherical boundary data on a final hypersurface, 
$\Sigma_{F}{\,}$,
separated from 
$\Sigma_{I}$
by a proper time 
$T$, 
as measured at spatial infinity. 
The initial data may be regarded as describing 
a nearly--spherical configuration of gravity and matter,
shortly before collapse to a black hole.
The weak final data may be regarded as describing a particular choice 
of very--late--time radiation.
The quantum amplitude to go from the prescribed initial data on
$\Sigma_{I}$ 
to the prescribed final data on
$\Sigma_{F}$
is essentially given by
$\exp\bigl(iS_{\rm cl}\bigr)$, 
where 
$S_{\rm cl}$ 
is the action of a solution of the classical field equations
which proceeds from the initial to the final data
(if such exists).
In particular, 
this is expected to hold in the case that the theory
is invariant under local supersymmetry [3].
In the present case,
this semi--classical expression for the quantum amplitude
will only be used in the weak--field limit, 
as above [43].
In fact, 
the above boundary--value problem,
of finding a classical solution to the wave--like 
(hyperbolic)   
equations of a field theory,
is not well--posed [4].
The natural cure, 
following Feynman [43],
is to rotate the time--interval 
$T$,
as measured at spatial infinity, 
slightly into the lower complex half--plane:
$$T
{\quad}
=
{\quad}
{\bigl\vert}T{\bigl\vert}{\,}e^{-{\,}i\delta}
{\quad},
\eqno(6.1)$$
\noindent
where 
$\delta{\;}{\,}
\bigl(0{\,}\leq{\,}\delta{\,}\ll{\,}{\pi}/2\bigr)$ 
is a real phase. 
One finds that 
$S_{\rm cl}$
has both its real and imaginary parts non--zero.
Hence, 
from the squared modulus of the quantum amplitude,
one arrives at a probability distribution 
for the weak--field radiation,
as measured on the final hypersurface 
$\Sigma_{F}$ 
[44].\par
\smallskip
\indent
Consider a particular configuration of the 
$f$ 
oscillators 
$N_{n}^{i}$ 
(assumed to be independent),
at each level 
$n$, 
associated with the boundary data on 
$\Sigma_{F}{\,}$. 
As above, 
we write:
$$N
{\quad}
=
{\quad}
N\bigl[N_{n}^{i}\bigr]
{\qquad},
{\qquad}
N_{n}
{\quad}
\equiv
{\quad}
\sum_{i=1}^{f}{\;}N_{n}^{i}
{\quad},
\eqno(6.2)$$
\noindent
The probability density has the form 
(cf. [44]):
$$P\bigl[\{N_{n}^{i}\}\bigr]
{\quad}
=
{\quad}
\Bigl[Z_{f}\bigl(\hat{g}^{2}\bigr)\Bigr]^{-1}{\;}
\exp\left(-{\;}\sum_{n=1}^{n_{\rm max}}{\;}\sum_{i=1}^{f}{\;}
\gamma_{n}{\;}N_{n}^{i}\right)
{\quad},
\eqno(6.3)$$
\noindent
where
$$\eqalignno{\gamma_{n}
{\quad}
&=
{\quad}
2{\pi}{\,}{\hat{g}}^{2}{\,}n{\;}
+{\;}2{\,}{\varepsilon}^{2}{\,}{\rho}_{n}
{\qquad}{\qquad},
{\qquad}{\qquad}
{\hat{g}}^{2}
{\quad}
=
{\quad}
{{\delta}\over{2}}
{\qquad},&(6.4)\cr
{\rho}_{n}
{\quad}
&=
{\quad}
{{2\pi}\over{\hat{T}}}{\;}{\;}
{\sum}_{k=-\infty}^{\infty}{\;}{\;}
\delta\biggl({\omega}_{n}{\;}-{\;}{{2{\pi}k}\over{\hat{T}}}\biggr)
{\quad}
=
{\quad}
{\sum}_{k=-\infty}^{\infty}{\;}{\;}
e^{ik{\omega}_{n}\hat{T}}
{\qquad},&(6.5)\cr}$$
\noindent
and the `partition function' is
$$Z_{f}\bigl(\hat{g}^{2}\bigr)
{\quad}
=
{\quad}
\sum_{\{N_{n}^{i}\}}{\,}\exp\left(-{\;}\sum_{n=1}^{n_{\rm max}}{\;}
\sum_{i=1}^{f}{\;}\gamma_{n}{\;}N_{n}^{i}\right)
{\quad}.
\eqno(6.6)$$ 
\noindent
The parameter
$\varepsilon$
is the small perturbation--expansion parameter 
given just before Eq.(1.1).
The 
$2{\pi}{\,}{\hat{g}}^{2}{\,}n$ 
term in 
$\gamma_{n}$ 
arises from the contribution to the classical action 
from the timelike boundary 
$\Sigma^{\infty}$ 
at spatial infinity, 
which joins 
$\Sigma_{I}$ 
to 
$\Sigma_{F}$ 
[32]. 
The 
${\varepsilon}^{2}$
term in 
$\gamma_{n}$ 
arises from the weak, 
non--spherical boundary data for integer--spin perturbations on 
$\Sigma_{F}$ 
[3].\par
\smallskip
\indent 
The 
$\{N_{n}^{i}\}$ 
sum in Eq.(6.6) is over all sequences of independent occupation numbers 
$N_{n}^{i}{\,}$, 
such that 
$n{\,}\geq{\,}1{\;},
{\,}1{\,}\leq{\,}i{\,}\leq{\,}f$. 
Despite the formal similarity of Eq.(6.6) 
to a thermodynamical partition function, 
one should view the parameter 
$2{\pi}{\,}{\hat{g}}^{2}$ 
as being conjugate to 
$N$, 
and not the energy. 
With Eqs.(6.1) and (6.5) in mind, 
this is analogous to a number--phase conjugacy in quantum mechanics, 
although 
$\delta$ 
is not an oscillator phase. 
A standard calculation gives
$$Z_{f}\bigl({\hat{g}}^{2}\bigr)
{\quad}
=
{\quad}
\prod_{n=1}^{n_{\rm max}}{\;}\Bigl(1{\,}-{\,}e^{-\gamma_{n}}\Bigr)^{-f}
{\quad},
\eqno(6.7)$$
\noindent
and from Eqs.(6.2,3),
we obtain the average
$$\bar{N}_{n}
{\quad}
=
{\quad}
{{f}\over{e^{\gamma_{n}}-1}}
{\quad}.
\eqno(6.8)$$
\indent  
A physical interpretation for 
${\hat{g}}^{2}$ 
(and hence for the phase 
$\delta$) 
can be obtained by evaluating 
$$N
{\quad}
=
{\quad}
\sum_{n=1}^{n_{\rm max}}{\;}n{\,}\bar{N}_{n}
\eqno(6.9)$$ 
\noindent
and using Eq.(6.8). 
Assuming that one can convert the sum to an integral 
(which is the case when
${\hat{g}}^{2}{\,}\ll{\,}1{\,}$), 
we find, 
from Eq.(6.3), 
that
$$2{\pi}{\,}{\hat{g}}^{2}
{\quad}
\simeq
{\quad}
N^{-{{1}\over{2}}}{\;}{\,}\sqrt{{{{\pi}^{2}f}\over{6}}}
{\quad}.
\eqno(6.10)$$
\noindent
Alternatively, 
let us assume that 
${\lambda}^{n}{\,}\gtrsim{\,}{\varepsilon}^{2}{\,}{\rho}_{n}{\,}$,
where 
$\lambda=2{\pi}{\,}{\hat{g}}^{2}{\,}$. 
Then,
$$P\bigl[\{N_{n}^{i}\}\bigr]
{\quad}
=
{\quad}
\Bigl[Z_{f}(\lambda)\Bigr]^{-1}{\;}e^{-{\,}\lambda N[\{N_{n}^{i}\}]}
{\quad}.
\eqno(6.11)$$
\noindent
Define
$$\rho(N)
{\quad}
=
{\quad}
\sum_{\{N_{n}^{i}\}}{\;}
\delta\Bigl[N{\,}-{\,}N\bigl(\{N_{n}^{i}\}\bigr)\Bigr]
{\quad}.
\eqno(6.12)$$
\noindent
Then,
$$Z_{f}(\lambda)
{\quad}
=
{\quad}
\int_{0}^{\infty}dN{\;}{\;}\rho(N){\;}{\,}e^{-\lambda N}
{\quad},
\eqno(6.13)$$
\noindent
which is just the Laplace transform of 
$\rho(N){\,}$. 
Therefore,
$$\hat{P}\bigl(N\bigr)
{\quad}
=
{\quad}
\Bigl[Z_{f}(\lambda)\Bigr]^{-1}{\;}{\,}
e^{-{\,}{\lambda}N}{\;}\rho\bigl(N\bigr)
\eqno(6.14)$$
\noindent
satisfies
$$\int_{0}^{\infty}dN{\;}\hat{P}\bigl(N\bigr)
{\quad}
=
{\quad}
1
{\quad}.
\eqno(6.15)$$
\noindent
Let us make the Ansatz
$$\rho\bigl(N\bigr)
{\quad}
=
{\quad}
c_{0}{\;}N^{-{{1}\over{2}}\beta}{\;}
\exp\Bigl(\alpha{\,}\sqrt{N}\Bigr)
{\quad},
\eqno(6.16)$$
\noindent
where 
$c_{0}{\,},\alpha$ 
and 
$\beta{\,}>{\,}0$ 
are all constants. 
Then, 
the probability density Eq.(6.14) is peaked around 
$N=N_{0}{\,}$, 
where
$$\sqrt{N_{0}}
{\quad}
=
{\quad}
{{1}\over{4\lambda}}{\;}{\,}
\Bigl[\alpha{\,}\pm{\,}\sqrt{\alpha^{2}-8\beta\lambda}\Bigr]
{\quad}
\eqno(6.17)$$
\noindent
If we further assume that 
${\alpha}^{2}{\,}\gg{\,}8\beta\lambda$ 
and that 
$N_{0}{\,}\gg{\,}1$, 
we obtain
$$\lambda
{\quad}
=
{\quad}
2{\pi}{\,}{\hat{g}}^{2}
{\quad}
\simeq
{\quad}
{{\alpha}\over{2{\,}\sqrt{N_{0}}}}
{\quad}.
\eqno(6.18)$$
\noindent
This agrees with Eq.(6.10), 
provided that  
$\alpha=2{\pi}\sqrt{f/6}$
and 
$N=N_{0}{\,}$. 
The inequality
${\alpha}^{2}{\,}\gg{\,}8\beta\lambda$ 
seems reasonable for large 
$f$. 
Therefore,
$$\rho\bigl(N\bigr)
{\quad}
=
{\quad}
c_{0}{\,}N^{-{{1}\over{2}}\beta}{\,}
\exp\left(2{\pi}\sqrt{{{fN}\over{6}}}\right)
{\quad}.
\eqno(6.19)$$
\noindent
The logarithm of Eq.(6.19) agrees with Eq.(5.6) to leading order. 
Eqs.(6.10,18) are analogous to the critical string coupling 
condition of Eq.(4.5).  
This suggests that the parameter 
$\hat{g}$ 
behaves like a string coupling parameter.\par
\smallskip
\indent 
There are further indications that the rotation angle 
$\delta$ 
is analogous to the 
(squared) 
string coupling. 
In the squeezed--states approach to black--hole evaporation [45], 
the squeeze parameter 
$r_{j}$ 
is related to the frequency 
$\omega_{j}$ 
and the black--hole mass 
$M$ 
through
$$\tanh{r_{j}}
{\quad}
=
{\quad}
\exp\left(-{\;}{{4{\pi}GM\omega_{j}}\over{c^{3}}}\right)
{\quad}.
\eqno(6.20)$$
\noindent
Since 
$<N_{j}>=\sinh^{2}(r_{j})$, 
this gives a thermal distribution 
$$<N_{j}>
{\quad}
=
{\quad}
{{1}\over{e^{\beta_{H}\omega_{j}}-1}}
{\quad},
\eqno(6.21)$$
\noindent
at the Hawking temperature 
$T_{H}
=\bigl(k_{B}{\,}\beta_{H}\bigr)^{-1}$.  
This spectrum is only valid in the phase for which the black hole
emits particles at low energy, 
(which, 
of course, 
corresponds to the majority of the lifetime of the black hole).  
For the final stages of evaporation, 
the energy spectrum does not have the canonical form, 
but a micro--canonical distribution [12]. 
In the approach of [44], 
the boundary--value problem of [3] for computing quantum amplitudes 
for wave--like massless perturbations at late times 
was related to the squeezed--states formalism. 
It was shown that the limit 
${\delta}{\,}\to{\,}0$
for the rotation angle in Eq.(6.1) 
corresponds to a highly--squeezed final state 
for the relic Hawking radiation, 
with a squeeze parameter 
$r_{j}$ 
given by
$$\tanh{r_{j}}
{\quad}
=
{\quad}
\exp\Bigl(-{\,}2{\,}{\omega}_{j}{\,}|T|{\,}\sin{\delta}\Bigr)
{\quad}.
\eqno(6.22)$$
\noindent
Eqs.(6.20) and (6.22) agree when
$${{hN\sin{\delta}}\over{Mc^{2}}}
{\quad}
=
{\quad}
{{4{\pi}GM}\over{c^{3}}}
{\quad}.
\eqno(6.23)$$
\noindent
Here, 
$N$ 
is given in Eq.(2.6). 
For small 
$\delta$, 
this gives Eq.(3.9), 
with 
$g$ 
replaced with 
$\hat{g}$.
\par
\smallskip
\indent 
In addition, 
Damour and Ruffini's approach to black--hole evaporation [46] 
involved complexifying the Schwarzschild mass 
$M$, 
in order to smooth out large oscillations 
in the outgoing particle wave--function 
as one approaches the future event horizon. 
Whether one continues 
$M$ 
into the lower or the upper half of the complex plane 
depends upon whether one wishes to describe positive-- 
or negative--energy states, 
respectively. 
In this procedure,
quantum states are smooth across the event horizon,
and describe a negative--energy component 
that falls towards the curvature singularity, 
together with a positive--energy component that can escape to infinity 
(depending on its energy). 
If, 
following this approach, 
one sets
$$M
{\quad}
=
{\quad}
{\bigl\vert}M{\bigl\vert}{\;}{\,}
\exp\bigl(-{\,}i\delta\bigr)
{\quad},
\eqno(6.24)$$
\noindent
then, 
for small 
$\delta$, 
one has 
$M
={\bigl\vert}M{\bigl\vert}
-\bigl(i{\,}\Gamma\bigr)/2{\,}$, 
where we define
$$\Gamma
{\quad}
=
{\quad}
4{\,}{\hat{g}}^{2}{\;}{\bigl\vert}M{\bigl\vert}
{\quad}.
\eqno(6.25)$$
\noindent
This expression bears a resemblance to the expression 
in Eq.(4.9) for the width or decay rate,  
taking 
$\hat{g}$ 
as in Eq.(6.5).\par
\smallskip
\indent 
This analysis suggests that the angle 
$\delta$,
by which the time--interval at spatial infinity, 
$T$,
is rotated into the complex plane, 
plays a r\^ole analogous to that of the string coupling 
$g$. 
Black--hole evaporation can therefore be viewed in the following way: 
Initially, 
$\delta$
is non--zero, 
and is given by its value in the Euclidean r\'egime. 
In the limit 
$\delta{\,}\to{\,}0_{+}$ 
(and, 
hence, 
for weak coupling 
$\hat{g}{\,}\to{\,}0$), 
one moves towards the physical, 
Lorentzian, 
r\'egime, 
and the black hole makes a transition to a `string--like' state, 
where it subsequently evaporates into massless particles.\par
\bigskip
\noindent 
{\bf 7. Conclusion}
\medskip
\indent 
In this paper, 
we have expressed the black--hole entropy 
{\it via} 
a counting of massless field configurations 
which have a fixed total energy 
on a post-evaporation space--like hypersurface.
We have also explored some of the analogies 
with the string/black--hole correspondence, 
incorporating some of the notions of black--hole evaporation 
in a complexified--time representation.\par

\parindent = 1 pt
\bigskip
{\bf References}
\medskip

\indent [1] J.D. Bekenstein, 
Phys. Rev. D {\bf 7} (1973) 2333.\par 

\indent [2] J.W. York,  
Phys. Rev. D {\bf 28} (1983) 2929.\par

\indent [3] A.N.St.J. Farley and P.D. D'Eath, 
Phys. Lett. B {\bf 601} (2004) 184 
(arXiv gr-qc/0407086);  
A.N.St.J. Farley and P.D. D'Eath, 
Ann. Phys. (N.Y.) {\bf 321} (2006) 1334; 
A.N.St.J. Farley and P.D. D'Eath, 
``Quantum amplitudes in black--hole evaporation. I:
Complex approach'',
(arXiv gr-qc/0510028);
A.N.St.J. Farley and P.D. D'Eath,  
``Quantum amplitudes in black--hole evaporation. II:
Spin--0 amplitude'',
(arXiv gr-qc/0510029).\par

\indent [4] P. R. Garabedian, 
{\it Partial Differential Equations}, (New York: Wiley) (1964).\par

\indent [5] S.W. Hawking, 
Phys. Rev. D {\bf 14} (1976) 2460.\par

\indent [6] A.N.St.J. Farley and P.D.D'Eath, 
in preparation.\par

\indent [7] S. Kalyana Rama,
``A stringy correspondence principle in cosmology'',
(arXiv hep-th/0603216).\par

\indent [8] P.C. Vaidya, 
Proc. Indian Acad. Sci. {\bf A33}, 264 (1951).\par

\indent [9] G. Gour, 
Phys. Rev. D {\bf 61} (2000) 124007 
(arXiv gr-qc/9906116); 
G. Gour, 
Phys. Rev. D {\bf 61} (2000) 021501 
(arXiv gr-qc/9907066).\par

\indent [10] A.N.St.J. Farley and P.D. D'Eath, 
Gen. Relativ. Gravit. {\bf 38} (2006) 425  
(arXiv gr-qc/0510043).\par

\indent [11] C. Rovelli,
Phys. Rev. Lett. {\bf 77} (1996) 3288 
(arXiv gr-qc/9603063).\par

\indent [12] B. Harms and Y. Leblanc,
Phys. Rev. D {\bf 46} (1992) 2334 
(arXiv hep-th/9205021).\par

\indent [13] J.D. Bekenstein, 
Lett. Nuovo Cim. {\bf 11} (1974) 467.\par

\indent [14] T. Rothman and P. Anninos, 
Phys. Rev. D {\bf 55} (1997) 1948 
(arXiv gr-qc/9612063); 
T. Rothman, 
Gen. Relativ. Gravit. {\bf 32} (2000) 1185 
(arXiv gr-qc/9906002).\par

\indent [15] N. Sasakura, 
Prog. Theor. Phys. {\bf 102} (1999) 169 
(arXiv hep-th/9903146).\par

\indent [16] J.L. Cardy, 
Nucl. Phys. B {\bf 270} (1986) 186.\par

\indent [17] G.T. Horowitz and J. Polchinski,
Phys. Rev. D {\bf 57} (1998) 2557 
(arXiv hep-th/9707170);
G.T. Horowitz and J. Polchinski,
Phys. Rev. D {\bf 55} (1997) 6189 
(arXiv hep-th/9612146);
G.T. Horowitz, 
``Quantum states of black holes'',
(arXiv gr-qc/9704072).\par

\indent [18] L. Susskind,  
``Some speculations about black hole entropy in string theory'',
(arXiv hep-th/9309145).\par

\indent [19] M.J. Bowick, L. Smolin and L.C.R. Wijewardhana,  
Gen. Relativ. Gravit. {\bf 19} (1987) 113;
M.J. Bowick, L. Smolin and L.C.R. Wijewardhana, 
Phys. Rev. Lett. {\bf 56} (1986) 424.\par

\indent [20] N.G. S\'anchez, 
Int. J. Mod. Phys. A {\bf 19}, 4173 (2004)
(arXiv hep-th/0312018);  
N. S\'anchez,  
``Black hole emission in string theory 
and the string phase of black holes'',
(arXiv hep-th/0106221).\par

\indent [21]  S.N. Solodukhin,  
Phys. Rev. D {\bf 57} (1998) 2410
(arXiv hep-th/9701106);  
R.K. Kaul,  
Phys. Rev. D {\bf 68}, 024026 (2003)
(arXiv hep-th/0302170).\par

\indent [22] T. Damour, 
Annalen Phys. {\bf 11} (2000) 1 
[Grav. Cosmol. Suppl. {\bf 6} 
(2000 ANPYA, 9, 267-277, 2000) 63] 
(arXiv hep-th/9912224);
T. Damour and G. Veneziano, 
Nucl. Phys. B {\bf 568} (2000) 93 
(arXiv hep-th/9907030).\par

\indent [23]  {~}S. {~}Dimopoulos {~~}and {~~}R. {~}Emparan,  
{~}Phys. {~}Lett. {~}B {~}{\bf 526} {~}(2002) {~}393
{~~}(arXiv {~~}hep-ph/0108060).\par

\indent [24]  K. Zalewski,  
Phys. Lett. B {\bf 207} (1988) 278.\par

\indent [25]  H.J. de Vega and N.G. S\'anchez,  
Phys. Rev. D {\bf 67} (2003) 125019.\par

\indent [26] J. Dai and J. Polchinski, 
Phys. Lett. B {\bf 220} (1989) 387;
R.B. Wilkinson, N. Turok and D. Mitchell,
Nucl. Phys. B {\bf 332}, 131 (1990).\par

\indent [27] D. Mitchell and N. Turok,
Nucl. Phys. B {\bf 294} (1987) 1138.\par

\indent [28] R.R. Khuri, 
Nucl. Phys. B {\bf 588} (2000) 253 
(arXiv hep-th/0006063);
R.R. Khuri,
Nucl. Phys. B {\bf 617} (2001) 365 
(arXiv hep-th/0107092);
R.R. Khuri, 
Mod. Phys. Lett. A {\bf 13} (1998) 1407
(arXiv gr-qc/9803095);
R.R. Khuri, 
Phys. Lett. B {\bf 470} (1999) 73
(arXiv hep-th/9910122); 
O. Bergman and C.B. Thorn,
Nucl. Phys. B {\bf 502} (1997) 309 
(arXiv hep-th/9702068);
C.B. Thorn, 
``Substructure of string'',
(arXiv hep-th/9607204).\par

\indent [29] D.V. Fursaev, 
``Can one understand black hole entropy
without knowing much about quantum gravity?'',
(arXiv gr-qc/0404038).\par

\indent [30] S. Das, A. Dasgupta and T. Sarkar, 
Phys. Rev. D {\bf 55} (1997) 7693 
(arXiv hep-th/9702075).\par

\indent [31] {~}A. {~}Krause, 
{~}``Black {~}holes, {~}space--filling {~}chains 
{~}and {~}random {~}walks'',
{~~}(arXiv {~~}hep-th/0312311).\par

\indent [32] G.W. Gibbons and S.W. Hawking,
Phys. Rev. D {\bf 15} (1977) 2752.\par

\indent [33] S. Das and V. Husain, 
Class. Quantum Grav. {\bf 20} (2003) 4387 
(arXiv hep-th/0303089).\par

\indent [34] D.N. Page, 
Phys. Rev. D {\bf 13} (1976) 198.\par

\indent [35] A. Ghosh, 
Phys. Lett. B {\bf 425} (1998) 269 
(arXiv hep-th/9801064).\par

\indent [36] L. Susskind, 
Phys. Rev. Lett. {\bf 71} (1993) 2367 
(arXiv hep-th/9307168);
L. Susskind, Phys. Rev. D 
{\bf 49} (1994) 6606 
(arXiv hep-th/9308139);
L. Susskind and J. Uglum, 
``Black holes, interactions, and strings'',
(arXiv hep-th/9410074).\par

\indent [37] R.K. Kaul and P. Majumdar,
Phys. Rev. Lett. {\bf 84} (2000) 5255 
(arXiv gr-qc/0002040).\par

\indent [38] S.S. More, 
``Higher order corrections to black hole entropy'', 
(arXiv gr-qc/0410071).\par

\indent [39] S. Carlip, 
Class. Quantum Grav. {\bf 17} (2000) 4175 
(arXiv gr-qc/0005017).\par

\indent [40] M.I. Park, 
``Testing holographic principle
from logarithmic and higher order corrections to black hole entropy'',
(arXiv hep-th/0402173).\par

\indent [41] S. Kalyana Rama, 
Phys. Lett. B {\bf 566} (2003) 152 
(arXiv hep-th/0304152).\par

\indent [42] J.l. Jing and M.L. Yan,
Phys. Rev. D {\bf 63} (2001) 024003 
(arXiv gr-qc/0005105).\par

\indent [43] R.P. Feynman and A.R. Hibbs,
{\it Quantum Mechanics and Path Integrals} 
(New York: McGraw-Hill) (1965).\par

\indent [44] A.N.St.J. Farley and P.D. D'Eath, 
Phys. Lett. B {\bf 634} (2006) 419.
(arXiv gr--qc/0603092);
Class. Quantum Grav. {\bf 24} (2007) 105.\par

\indent [45] L.P. Grishchuk and Y.V. Sidorov, 
Phys. Rev. D {\bf 42} (1990) 3413.\par

\indent [46] T. Damour and R. Ruffini, 
Phys. Rev. D {\bf 14} (1976) 332.\par

\end